\documentstyle[12pt,epsf]{article}
\begin{document}
\voffset -2cm
\newcommand{\be}{\begin{equation}}
\newcommand{\bea}{\begin{eqnarray}}
\newcommand{\ee}{\end{equation}}
\newcommand{\eea}{\end{eqnarray}}
\newcommand{\lb}{\label}
\newcommand{\nl}{\newline}
\newcommand{\ra}{\rightarrow}
\newcommand{\nn}{\nonumber}
\newcommand{\al}{\alpha}
\newcommand{\de}{\delta}
\newcommand{\eps}{\epsilon}
\newcommand{\ga}{\gamma}
\newcommand{\la}{\lambda}
\newcommand{\om}{\omega}
\newcommand{\ns}{\normalsize}
\newcommand{\ph}{\phi}
\newcommand{\pr}{\prime}
\newcommand{\sld}{/\hspace{-0.55em}\partial}
\newcommand{\slk}{/\hspace{-0.55em}k}
\newcommand{\slp}{/\hspace{-0.58em}p}
\newcommand{\slq}{/\hspace{-0.5em}q}
\newcommand{\mi}{\mbox{i}}
\newcommand{\mt}{m_{top}}
\renewcommand{\thesection}{\arabic{section}.}
\renewcommand{\thefootnote}{\fnsymbol{footnote}}
\begin{titlepage}
\begin{flushleft}
{\bf DESY 93-111 \\ LMU-04/93} \\ {\em August 1993}
\end{flushleft}
\vspace{1.5cm}
\begin{center}
{\bf EXPERIMENTAL CONSTRAINTS ON THE SCALE OF NEW PHYSICS
\\ IN TOP CONDENSATE MODELS}
\end{center}
\vspace{1cm}
\begin{center}{\bf Ralf B\"{O}NISCH$^1$
 and Arnd LEIKE$^2$}

\vspace{1cm}
$^{1}$Ludwig-Maximilians-Universit\"{a}t M\"{u}nchen, Sektion Physik  \\
80333 M\"{u}nchen, Germany \\
\vspace{0.2cm}
$^{2}$DESY - Institut f\"ur Hochenergiephysik, \\
Platanenallee 6, 15735 Zeuthen, Germany
\end{center}
\vspace{1cm}
\begin{abstract}
We obtain mass limits on the extra neutral gauge boson which is predicted in a
model with hidden gauge symmetry and
dynamical breaking of the electroweak symmetry by a top quark condensate.
For typical model assumptions, present LEP data exclude masses below 3\,TeV.
With LEP200 or an electron-positron collider of a c.m. energy
of 500 GeV masses below 15\,TeV or 50\,TeV could be excluded, respectively.
Such high mass limits allow the calculation of the observables used in the
analysis in the context of the Nambu-Jona-Lasinio model.
\end{abstract}
\vspace{2cm}
\end{titlepage}
\renewcommand{\thefootnote}{\fnsymbol{footnote}}
%
\section{Introduction}
It is well known that precision experiments at $e^+e^-$ colliders
can give indirect limits on masses of extra neutral gauge bosons.
However, the best mass limits on weakly coupled gauge bosons are usually
set by direct production at hadron colliders.
In this letter, we consider a model which contains an extra, strongly coupled
neutral gauge boson originating in top condensate models. We derive
mass limits from present and future $e^+e^-$ colliders
and find that here they
are superior in comparison with hadron colliders.

The large mass of the top quark motivates a specific dynamical
breakdown of the electroweak symmetry by the formation of a top
quark condensate \cite{claro}.
The Top Mode Standard Model (TMSM) possesses
scalar four fermion interactions of Nambu-Jona-Lasinio type instead of
elementary scalar Higgs-fields, \cite{nambu} - \cite{bhl}.
In its minimal form \cite{bhl} it hardly matches the experimentally
allowed top mass window \cite{rho}.
Moreover, the TMSM has been found unsatisfactory for theoretical reasons
concerning predictability beyond the Standard Model (SM) \cite{critics,boe}.
The naturalness requirement demands the scale of next physics beyond the
SM to be in the TeV range.
Here, far below a GUT scale,
one does not expect to already leave the framework of quantum
gauge field theories and the TMSM has never been meant to solve the fundamental
mass problem.
The simple possibility to assume a well-defined gauge theory as the origin
of the NJL-type model \cite{boe} attempts to find the structure beyond the SM
in using the fact of the large top mass.
A variety of gauge extensions to $SU(2)_L \times U(1)_Y \times G$
was proposed on these grounds \cite{boe2,models,boekneur}.
The breaking of non-standard symmetries is thereby
often described by scalar fields and this
does not disturb the concept of looking for structures at the TeV scale.
 The value of the new scale, related to the mass of new bosons,
is essential for theoretical predictions of the models.

In this letter, we derive lower bounds for the mass of
non-standard gauge bosons within the Hidden Symmetry Model \cite{boekneur}
from $e^+e^-$ data.

In section 2 we briefly introduce the features of the Hidden Symmetry Model
needed for the analysis of experimental data. In section 3 we
describe the analysis and discuss the results. We conclude in section 4.
%
\section{The Hidden Symmetry Model}
The extension to the hidden gauge group $SU(2)_V$
has been discussed in
\cite{bmw} and recently in  \cite{ks}.
It was applied to the top condensate mechanism in
\cite{boe,boekneur}.

Starting point is the symmetry $SU(2)_L \times U(1)_{Y^\pr} \times
SU(2)_V$. [As none of the features of the model, which make
it special for top quark condensation, depends on the dimension
of the hidden group, it could as well be any hidden $SU(N)$ including $U(1)$.]
$U(1)_{Y^\pr}$ differs from the standard $U(1)_Y$ only by the definition
of the gauge coupling constant.
$SU(2)_V$ is a hidden symmetry, i.e.
fermions are standard: they do not possess new degrees of freedom and
transform as singlets.
These local symmetries are broken down to electromagnetism in two steps:
\be  \begin{array} {lcr}
SU(2)_L \times U(1)_{Y^\pr}  \times SU(2)_V
& \stackrel{spontaneously\hspace{.1cm} by\hspace{.1cm} scalars}
{\longrightarrow}& \\
SU(2)_L \times U(1)_{Y}
&\stackrel{dynamically\hspace{.1cm} by\hspace{.1cm} \langle \bar{t}t \rangle}
{\longrightarrow} &
U(1)_{em}
\lb{breakpattern}\end{array}
\ee
The first step looks very much like the spontaneous symmetry breaking
in the SM, but
here SU(2)$_V \times $ U(1)$_{Y^\pr}$ is broken down to U(1)$_Y$.
Diagonalization causes a mixing of primordial U(1)$_Y^\pr$
and SU(2)$_V$ fields
$\bar{B}$ and $\bar{V}$ to a massless $B$ and a massive $v^0$ in the neutral
sector. The mixing angle $\xi$ is given by
\be \sin \xi = \frac{g^\pr}
{\sqrt{ g^{\pr \pr 2} + g^{\pr 2} }}.  \ee
Between the new scale and the Fermi scale, i.e. in the
$SU(2)_L \times U(1)_{Y} $ symmetric phase,
the interaction of neutral gauge bosons with fermions
is given as \cite{boekneur}
\begin{eqnarray}
{\cal L}_{0} = g J_i^\mu W_{\mu i} + g'\cos\xi J_Y^\mu B_\mu
- g'\sin\xi J_Y^\mu v^0_\mu .
\lb{ww}
\end{eqnarray}
$J_i^\mu$ and  $J_Y^\mu$ are exactly the SM-isospin and hypercharge currents.
\ \ $g'\cos\xi$\ \ is restricted to have the value of the standard hypercharge
gauge coupling constant. Charged heavy bosons $v^\pm$ do not couple to standard
fermions.

A strong coupling
\begin{equation}
(g'\sin\xi )^2\frac{2}{18} > \frac{8\pi ^2}{N} = G_c
\lb{gc}\end{equation}
causes the condensation of $\bar{t}t$. Down type quarks do not form
condensates because of a repulsive interaction by means of the hypercharge
quantum numbers. In flavor space, the mass matrix has rank one, so that
finally only the top achieves a mass as wanted \cite{boe2}.
The present model is special in not assuming a heavy top quark.
Instead, the only source of explicit $SU(2)_R$ violation in the SM,
the hypercharge, is used to produce a heavy top and a light bottom quark.

The strongly interacting new sector, eq. (\ref{gc}),
cannot be treated with perturbation theory.
It is handled by ladder
approximation and an expansion in $p^2/M^2_v$,
where $M_v$ is the mass of the new neutral gauge boson and
$p$ is the typical energy scale, i.e. $p^2=M_Z^2$ for LEP I.
Both expansions must be
motivated. The ladder diagrams are dominant in summing over an appropriate
large number $N$ of fermionic degrees of freedom. There is a very close
connection of the large $N$ expansion to the resulting mass matrix \cite{diss}.
In the Hidden Symmetry Model, $N$ is hidden and understood to carry information
of the inner structure of fermions. (It is not related to the size of the
hidden gauge group.) The restriction to first order in $p^2/M^2_v$ allows
the analytical calculation of bound states, because ladder diagrams sum as
geometric series in this limit. This approximation can always be used
as will be shown by experimental limits on $M_v$.
Deviations from these approximations would modify the relation of $\mt$ to the
$\rho$-parameter, as discussed in the minimal TMSM \cite{bhl,paschos}.
Other sources of deviations from those predictions are vector resonances, which
should additionally appear in the gauge models.

A typical model assumption is $N=3$. The corresponding critical coupling is
\begin{equation}
g'\sin\xi > \nu 2 \pi \sqrt{6}  = 2 \pi \sqrt{6},\ \ \ \nu=\sqrt{3/N}.
\end{equation}
%
\section{Analysis of $e^+e^-$ Data}
We now search for mass limits to the extra neutral gauge boson $V$ from
present and future electron-positron colliders.
The remarkable agreement of present experiments with the SM fix its
parameters with high precision and leads
to definite predictions at higher energies.
A direct production of extra neutral gauge bosons in $e^+e^-$ collisions is
very unlikely because of present mass limits from hadron colliders.
Although, it can be observed indirectly, if
one observable deviates from the SM prediction more than
the expected experimental uncertainty.

In our analysis, we take into account statistical and systematic errors as
well as radiative corrections.
We assume an integrated luminosity of
$L_{int}=20 fb^{-1}$ at $\sqrt{s} = 500GeV$ for a linear electron positron
collider (Linac) and $L_{int}=0.5 fb^{-1}$ at $\sqrt{s} = 190GeV$ for LEP200,
which corresponds roughly to one year of running time.
Dominant systematic errors at the Linac (LEP200) are due to luminosity
uncertainty and the errors of lepton and hadron energy measurements of
1\% (0.5\%), 0.5\% (0.5\%) and 1\% (1\%), respectively \cite{EE500}.

The necessary QED corrections are taken into account including a cut
$\Delta$ on the photon energy $E_\gamma /E_{beam} < \Delta = 0.7$. $\Delta$
is needed to remove the radiative tail and to
suppress the background \cite{EE500}.

We use an extended version of the program {\tt ZCAMEL} \cite{EE500} for the
calculation of observables (cross sections and asymmetries). It includes the
full $O(\alpha)$ QED corrections in theories with extra neutral gauge bosons
and soft photon exponentiation \cite{zecorr}.
We include the following observables in our analysis
\begin{eqnarray}
\label{genobs}
\sigma_t(\bar{l}l),\ A_{FB}(\bar{l}l),
\ R=\sigma_t(\bar{l}l)/\sigma_t(\bar{q}q),\
A_{LR}(\bar{l}l),\\ \nonumber
\ A_{LR}(\bar{q}q),\
 \sigma_t(\bar{b}b),\ A_{FB}(\bar{b}b),
\ A_{LR}(\bar{b}b),\ P_\tau,\ P_\tau^{FB},
\end{eqnarray}
where $\bar{l}l$ ($\bar{q}q$) refer to the production of leptons (5 quarks),
respectively, and $P_\tau$ and $P_\tau^{FB}$ are $\tau$ polarization
asymmetries.

We found that for LEP200 and for the Linac the total cross section of lepton
production gives the best mass limits to the $V$-boson mass.
For the special case $N=3$, we obtain
\begin{eqnarray}
\label{lep2}
M_V & > & 15\, TeV, \mbox{\ \ 95\% c.l.\ \ for LEP200},\\ \nonumber
M_V & > & 50\, TeV, \mbox{\ \ 95\% c.l.\ \ for the Linac.}
\end{eqnarray}
Beam polarization can improve these bounds due to
the left-right asymmetry of hadron production $A_{LR}(\bar{q}q)$.
For $N\ne 3$, the limits are shown in Fig.~1 and Fig.~2.

The mass limits from LEP~I data are different. In a full analysis one has to
constrain the Standard Model parameters, the mass of the extra neutral gauge
boson and its mixing with the $Z$-boson simultaneously \cite{zefitsr}.
LEP data constrain mainly the mixing.
In the Hidden Symmetry Model there is no mixing between
the $Z$ and $V$ by definition because they are mass eigenstates.

A full analysis of LEP data
in the Hidden Symmetry Model has not yet been done.
However, we observed that the leptonic forward-backward asymmetry at the
$Z$-peak $A_{FB}(\bar{l}l)$ is much more sensitive to $M_V$ than the other
observables at LEP.  We obtained an approximate mass limit to the $V$ by
the following procedure: We took the observables
\begin{equation}
P_\tau,\ P_\tau^{FB},\ A_{FB}(\bar{b}b),\ A_{FB}(\bar{c}c),\ A_{FB}(\bar{q}q),
\end{equation}
measured at the $Z$-peak \cite{moriond} and their errors and calculated a
prediction for $A_{FB}(\bar{l}l)$. It is, of course, consistent with
the measured value of $A_{FB}(\bar{l}l)$. Demanding that an additional neutral
gauge boson $V$ should not spoil
this consistency leads to the mass limits shown in Fig.~2. For $N=3$, we get
\begin{equation}
\label{lep1}
M_V > 3\, TeV, \mbox{\ \ 95\% c.l.\ \ for LEP~I}.
\end{equation}
%
\ \vspace{1cm}\\
\begin{minipage}[t]{7.8cm}{
\begin{center}
\hspace{-1.7cm}
\mbox{
\epsfysize=7.0cm
\epsffile[0 0 500 500]{zp3lep12.ps}
}
\end{center}
\noindent
{\small\bf Fig.~1: }{\small\it
The LEP mass limits for the additional neutral $V$-boson as function of the
coupling strength.
}
}\end{minipage}
\hspace{0.5cm}
\begin{minipage}[t]{7.8cm} {
\begin{center}
\hspace{-1.7cm}
\mbox{
\epsfysize=7.0cm
\epsffile[0 0 500 500]{zp3nlc.ps}
}
\end{center}
\noindent
{\small\bf Fig.~2: }{\small\it
The 500 GeV collider mass limits for the additional neutral $V$-boson as
function of the coupling strength.
}
}\end{minipage}
\vspace*{0.5cm}
%

The mass limits (\ref{lep2}), (\ref{lep1}) have a linear dependence on the
coupling constant $g'\sin\xi$, see Fig.~1, Fig.~2.
This is due to the large $V-$fermion coupling enabling us to be sensitive to
$V$-bosons, which are much heavier than the c.m. energy.
Thus, a deviation of any observable from the
SM prediction depends only on the ratio $g'\sin\xi / M_{V}$ and not on
$g'\sin\xi$ and $M_{V}$ separately:
\begin{equation}
\mbox{SM  } - \mbox{  Hidd. Symm.}
= -\mbox{const.} \frac{(g'\sin\xi)^2}{s-M_{V}^2+i\Gamma_{V} M_{V}}
\approx \mbox{const.}\left(\frac{g'\sin\xi}{M_{V}}\right)^2.
\end{equation}

The obtained limits on the $V$-mass constrain the mass scale of the hidden
symmetry breaking to be much larger than the mass scale of the electroweak
symmetry breaking, i.e. for all considered electron-positron colliders
$s/M_V^2 \ll 1$.
Hence, higher orders of the
parameter $s/M_V^2$ can be neglected and observables can be calculated
in the Hidden Symmetry Model.
Here we obtain consistency with our preliminary assumption
needed for the experimental analysis.
The case of a light $V$-boson can be also excluded because it would have
shown up in hadron colliders
or by an inconsistency of the observables at LEP with the SM due to its
large couplings to all fermions.

Finally, we mention that
we assumed the absence of unknown non-decoupling resonance effects
of the hidden strong interaction in the observables (\ref{genobs}).
\section{Conclusion}
Compared to hadron colliders, electron-positron colliders are superior in
setting mass limits to strongly coupled extra neutral gauge
bosons $V$. For the Hidden Symmetry Model considered
here, these bosons must be heavier than 3\,TeV to be consistent with present
LEP data. LEP200 or an $e^+e^-$
collider with a c.m. energy of 500\,GeV would improve these limits to 15\,TeV
or 50\,TeV.

As a consequence, higher orders in $p^2/M_V^2$ can be neglected, so that
ladder diagrams can be summed as usually done in the NJL model and
reliable predictions can be derived.
\vspace{1cm}\\ {\Large\bf Acknowledgement\hfill}\vspace{0.5cm}\\
We would like to thank T. Riemann for a careful reading of
the manuscript. \vspace{1cm}\\

\end{document}